\documentclass{ws-ijqi}

\newcommand{\ket}[1]{\left\vert#1\right\rangle}

\newcommand{\minivalmed}[1]{\langle#1\rangle}

\usepackage{epstopdf}
\usepackage{color}

\begin{document}

\markboth{C. Di Franco, M. Paternostro, and M. S. Kim}
{Bypassing state initialization in Hamiltonian tomography of spin chains}

\catchline{}{}{}{}{}

\title{BYPASSING STATE INITIALIZATION IN HAMILTONIAN TOMOGRAPHY ON SPIN-CHAINS}

\author{C. DI FRANCO}

\address{Department of Physics, University College Cork, Cork, Republic of Ireland\\
c.difranco@qub.ac.uk}

\author{M. PATERNOSTRO, M. S. KIM}

\address{School of Mathematics and Physics, Queen's University, Belfast BT7 1NN, United Kingdom}

\maketitle


\begin{abstract}
We provide an extensive discussion on a scheme for Hamiltonian tomography of a spin-chain model that does not require state initialization [Phys. Rev. Lett. 102, 187203 (2009)]. The method has spurred the attention of the physics community interested in indirect acquisition of information on the dynamics of quantum many-body systems and represents a genuine instance of a control-limited quantum protocol. 
\end{abstract}


\section{Introduction}

Recently, there has been a significant theoretical interest in the formulation of experimental-friendly criteria for the acquisition of information on the details of a given interaction model within a quantum many-body system.\cite{iceberg,burgarth,burgarthNJP,wm,stevemauro,huelga} The motivations behind this are manifold. On one hand, the ability to manipulate and control quantum systems made out of a few elementary constituents (such as multi-photon states, multi-spin states in liquid nuclear magnetic resonance or cavity-implanted superconducting devices) has greatly improved in the last ten years, making the possibility to explore the dynamics of quantum few-body systems a reality. Clearly, the effects to test with the help of one of such systems would heavily depend on the arrangement of the correct and wanted form of interaction. On the other hand, many of the most advanced protocols for  quantum information processing and communication rely on a fine pattern of coupling strengths across a lattice of interacting particles. This is important, for instance, when the task is to  transfer a quantum state across a spin chain.\cite{sougato} It is thus very interesting to find a way to {\it certify} with a reasonable degree of confidence that the set of coupling strengths within a given experimental device is the one required by the associated theoretical protocol.

From a practical viewpoint, it would be highly desirable to test if this is the case {\it before} running a protocol and to keep the invasiveness of such a test to minimum levels. A few important theoretical steps have been performed in this direction. In Ref.~\refcite{burgarth}, Burgarth {\it et al.} have proposed a scheme for the estimation of the parameters in a quantum spin-chain that is based on a clever application of the {\it inverse eigenvalues problem}, which is very well-known from classical inverse elastic theory.\cite{gladwell} Along similar lines, in Ref.~\refcite{iceberg} we have proposed a scheme for ``Hamiltonian tomography", {\it i.e.} the determination of the coupling parameters in a chain of interacting spins, which requires only time-resolved measurements over a single particle, simple data post-processing and no initialization or prior knowledge on the state of the whole system. 

Our analysis takes advantage of an information flux approach\cite{informationflux} that turns out to be particularly well-suited for the class of Hamiltonians to be probed by our tomographic scheme. In fact, our method allows us to gather very important knowledge on the temporal behavior of the whole system simply by looking at the dynamics of a single element. From that, the parameters present in the Hamiltonian can be inferred. The protocol is efficient even when the spin-chain is affected by Markovian dissipative and dephasing channels and can be generalized to Hamiltonian models that do not preserve the number of spin-excitations ({\it i.e.} interactions that do not commute with the total spin operator of the system). Very recently, Burgarth {\it et al.} have extended the basic idea of our Hamiltonian tomography method to general quadratic models, including the effects of transverse local magnetic fields and Ising-like terms.\cite{burgarth2} Analogously, Wie\'sniak and Markiewicz have discussed Hamiltonian diagnostic tools that do not rely on state preparation.\cite{wm} In this paper we provide a detailed account of the basic working principle behind the scheme discussed in Ref.~\refcite{iceberg}. The motivation behind Hamiltonian tomography, common to all the protocols put forward so far, is to provide a reliable way to fully characterize a many-body coupling model. As such, it can be seen as complementing and extending schemes that are already very well-known and successfully implemented, such as quantum process tomography,\cite{nielsenchuang} quantum state tomography\cite{QST} and detector tomography.\cite{QDT} We hope that the non-demanding nature of the class of Hamiltonian tomography protocols suggested so far will soon spur the attention of the experimental community interested in many-body dynamics.

The remained of the paper is organized as follows. In Sec.~\ref{tomography}, we introduce the Hamiltonian tomography protocol for various classes of interactions. In Sec.~\ref{simulation}, we use the scheme to determine the parameters in an engineered chain of eight spins. This serves as an illustration of the efficiency and working features of our protocol. Finally, we summarize our results in Sec.~\ref{remarks}.

\section{Hamiltonian tomography for different spin-chain models}
\label{tomography}
\begin{figure}[t]
\centerline{\psfig{figure=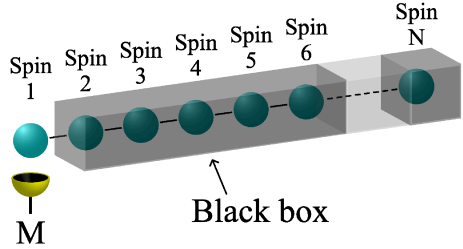,width=6cm}}
\caption{Sketch of the scheme for Hamiltonian tomography without state initialization, where $M$ is the measurement performed on spin $1$ and other spins are not accessible.}
\label{icebergscheme}
\end{figure}

The Hamiltonian tomography protocol proposed in Ref.~\refcite{iceberg} can be applied to various spin-chain models. A sketch of the scheme is presented in Fig.~\ref{icebergscheme}.
The general scenario behind Hamiltonian tomography considers a chain of $N$ spins mutually coupled according to a Hamiltonian model $\hat{\cal H}$. Of them, only the first spin, labelled hereafter as $1$, is fully accessible and can be measured in any basis, the remaining sites being out of reach. Our task is to determine the coupling strengths entering $\hat{\cal H}$ by means of the information that we can extract from the measurements performed on $1$ only. 

In Ref.~\refcite{iceberg}, we have discussed how the method can be used to infer the coupling parameters in $X\!X$-like interactions described by the Hamiltonian 
\begin{equation}
\label{XXmodel}
\hat{\cal H}_1=\sum^{N-1}_{i=1}J_{i}(\hat{X}_{i}\hat{X}_{i+1}+\hat{Y}_{i}\hat{Y}_{i+1}),
\end{equation}
as well as the more general $X\!Y$-like models
\begin{equation}
\label{XYmodel}
\hat{\cal H}_2=\sum^{N-1}_{i=1}(J_{X,i}\hat{X}_{i}\hat{X}_{i+1}+J_{Y,i}\hat{Y}_{i}\hat{Y}_{i+1}).
\end{equation}

In the former (latter) case, $J_i>0$ ($J_{\sigma,i}>0$, with $\sigma=X,Y$) is the anti-ferromagnetic coupling strength of the pairwise interaction between qubits $i$ and $i+1$. The operators $\hat{X}_i$, $\hat{Y}_i$ and $\hat{Z}_i$ denote the $x$, $y$ and $z$-Pauli matrix of spin $i$, respectively. The condition of anti-ferromagnetic coupling can be dropped if the signs of all the parameters are known. It is also possible to extend the method to another class of Hamiltonians. In fact, using the results of Ref.~\refcite{dimitris}, one can apply the scheme to determine the coupling strengths in the Ising-like model with a local transverse magnetic field given by
\begin{equation}
\label{dimitrismodel}
\hat{\cal H}_3=\sum^{N-1}_{i=1}J_{Z,i}\hat{Z}_{i}\hat{Z}_{i+1}+\sum^N_{i=1}B_i\hat{X}_i.
\end{equation}
Here, $J_{Z,i}>0$ is the anti-ferromagnetic coupling strength of the pairwise interaction between qubits $i$ and $i+1$ and $B_i$ is the strength of the coupling of spin $i$ to the local magnetic field affecting it. 

The key point of the scheme is that, for all the models described above, the time-dependent expectation values of simple single-spin operators contain information about the full set of coupling parameters. For instance, for the interaction model in Eq.~(\ref{XXmodel}), if the initial state of spin $1$ is
\begin{equation}
\ket{\pm_x}_1=\frac{1}{\sqrt{2}}(\ket{0}_{1}\pm\ket{1}_{1}),
\end{equation}
we obtain
\begin{equation}
\minivalmed{\hat{X}_1}(t)=\pm\alpha_1(t),
\end{equation}
with $\alpha_1(t)$ a function that depends on all the $J_i$'s [see Eqs.~(\ref{alphaexpansion}) and (\ref{coeff}) for a fully formal expression of such function]. On the other hand, for the model in Eq.~(\ref{XYmodel}), we need to consider the expectation values of two operators. The first one is again $\minivalmed{\hat{X}_1}(t)$, which provides the function $\alpha_1(t)$ in a way completely analogous to what discussed above. As for the other one, we need to initialize spin $1$ in an eigenstate of $\hat{Y}_1$
\begin{equation}
\ket{\pm_y}_1=\frac{1}{\sqrt{2}}(\ket{0}_{1}\pm i\ket{1}_{1}),
\end{equation}
so that we have
\begin{equation}
\minivalmed{\hat{Y}_1}(t)=\pm\beta_1(t).
\end{equation}
Finally, for the Hamiltonian in Eq.~(\ref{dimitrismodel}), we need to initialize spin $1$ in one of the eigenstates of $\hat{Z}_1$, which we label $\{\ket{0}_1,\ket{1}_1\}$, so as to obtain
\begin{equation}
\minivalmed{\hat{Z}_1}(t)=\pm\alpha_1(t). 
\end{equation}
Notwithstanding the difference among the various models, the tomographic scheme required in order to infer the coupling parameters is the same. Such similarity holds in virtue of the fact that the evolution of the operators associated with spin $1$ depends on the operators of all the spins in the same way. This can be very effectively captured in a graphical way by using the information flux viewpoint put forward in Refs.~\refcite{informationflux} or by noticing that, adopting the language used in Ref.~\refcite{burgarth2}, spin $1$ is {\it infected} by the rest of the chain in the same way, regardless of the details of the Hamiltonian model. 

\section{Hamiltonian tomography at work}
\label{simulation}
In order to show the efficiency and working features of the scheme, here we apply it to the case of a seven-coefficient problem. We have generated the set of randomly picked positive numbers $\{1.40,1.48,1.06,0.80,1.36,0.97,0.66\}$. Depending on the model to be probed, they can be taken to correspond to specific parameters in the associated Hamiltonian. For instance, for the interaction in Eq.~(\ref{XXmodel}), they can simply correspond to the full set of $J_i$'s. When Eq.~(\ref{XYmodel}) is considered, they can be taken as the elements of a set of parameters constructed by taking $J_{X,i}$ and $J_{Y,i}$ alternatively (for instance, we can identify them with $\{J_{X,1},J_{Y,2},J_{X,3},J_{Y,4},J_{X,5},J_{Y,6},J_{X,7}\}$). Finally, the random coefficients may correspond to the set of parameters $\{B_1,J_{Z,1},B_2,J_{Z,2},B_3,J_{Z,3},B_4\}$ for a 4-spin chain evolving under the Hamiltonian in Eq.~(\ref{dimitrismodel}).

For the sake of definiteness, we now describe the tomographic method in relation to the Hamiltonian in Eq.~(\ref{XXmodel}). It should be clear, though, that the corresponding results will hold for the other Hamiltonian settings, with due adjustments. Our approach is as follows: we determine $\minivalmed{\hat{X}_1}(t)$ at a few instants of time by simulating the evolution of the chain. This provides a discrete sampling of function $\alpha_1(t)$ which, as we said, contains full information on the coupling coefficients $J_i$'s. In order to extract the values of such parameters, we shall fit the data points with a proper functional form. We have found that the trial function
\begin{equation}
\label{try}
\alpha_1^{\rm{(try)}}(t)=\sum_{i=1}^{N/2}{\cal A}_i\cos(\omega_it)
\end{equation}
is in excellent agreement with the behavior of the simulated data. This allows us to determine the values of the amplitudes ${\cal A}_i$'s and frequencies $\omega_i$'s. We now use the fact that $\alpha_1(t)$ can be written as~\cite{informationflux}
\begin{equation}
\label{alphaexpansion}
\alpha_1(t)=\sum_{l=0}^{\infty}\frac{(2t)^l}{l!}\delta_1^{(l)}
\end{equation}
with $\delta_1^{(l)}$  given by the recurrence formula
\begin{equation}
\label{coeff}
\delta_j^{(l)}=(-1)^{j}[J_{j-1}\delta_{j-1}^{(l-1)}+J_{j}\delta_{j+1}^{(l-1)}],
\end{equation}
where we have $J_0=J_N=0$ and the initial conditions $\delta_j^{(0)}=0$ ($1$) for $j\ne 1$ ($j=1$).

\begin{figure}[t]
\centerline{\psfig{figure=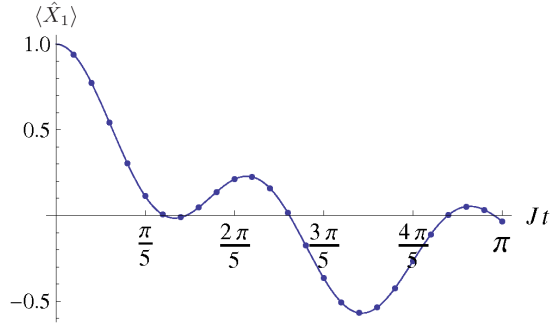,width=7.25cm}}
\caption{Simulated dynamics of $\minivalmed{\hat{X}_1}(t)$ under the action of $\hat{\cal H}$ with $N=8$ and $\{J_i/J\}=\{1.40,1.48,1.06,0.80,1.36,0.97,0.66\}
$, sampled at steps of $Jt=\pi/25$. The corresponding fit, performed using the trial function in Eq.~(\ref{try}), is also shown.}
\label{plot1}
\end{figure}

Both Eqs.~(\ref{try}) and (\ref{alphaexpansion}) are then expanded in Taylor series. By equating term by term the two series, one ends up with a linear system of algebraic equations whose solution gives the set of coupling parameters. Let us now illustrate the quantitative performance of the scheme:  in Fig.~\ref{plot1}, we show the dynamics of $\minivalmed{\hat{X}_1}(t)$ under the action of the Hamiltonian $\hat{\cal H}_1$ constructed by means of the random set of coefficients given above. The expectation value $\minivalmed{\hat{X}_1}(t)$ is sampled at steps $Jt=\pi/25$ (the corresponding values are shown as dots in the plot).
By fitting such data with Eq.~(\ref{try}), we have obtained the following values for ${\cal A}_i$'s and $\omega_i$'s
\begin{equation}
\begin{split}
{\cal A}_1=0.0921,&\hskip0.25cm\omega_1=3.5929,\\
{\cal A}_2=0.2748,&\hskip0.25cm\omega_2=4.4941,\\
{\cal A}_3=0.3155,&\hskip0.25cm\omega_3=0.7821,\\
{\cal A}_4=0.3176,&\hskip0.25cm\omega_4=1.6909.
\end{split}
\end{equation}
Having such $N$ values (that satisfy the constraint $\sum_{i=1}^{N/2}{\cal A}_i=1$), we need $N-1$ equations in order to find the $N-1$ parameters of the Hamiltonian under investigation. The first $N-1$ non-zero terms in the Taylor expansion of $\alpha_1^{\rm{(try)}}$ can be written as
\begin{equation}
\label{eta}
\eta_j=(-1)^j\frac{1}{(2j)!}\sum_{i=1}^{N/2}{\cal A}_i^j\omega_i^{2j}.
\end{equation}
We need to find the corresponding terms $\mu_j$ in the expansion of $\alpha_1(t)$ as in Eq.~(\ref{alphaexpansion}). Here we report only the first $4$ terms (the other ones being rather cumbersome)
\begin{equation}
\label{mu}
\begin{split}
\mu_1=&-2J_1^2,\\
\mu_2=&\frac{2}{3}(J_1^4+J_1^2J_2^2),\\
\mu_3=&-\frac{4}{45}J_1^2[(J_1^2+J_2^2)^2+J_2^2J_3^2],\\
\mu_4=&\frac{2}{315}J_1^2\{J_1^6+3J_1^4J_2^2+J_1^2(3J_2^4+2J_2^2J_3^2)+J_2^2[(J_2^2+J_3^2)^2+J_3^2J_4^2]\}.
\end{split}
\end{equation}
By solving the system of equations $\eta_j=\mu_j$ $(j=1,...,7)$, one finds the set of estimated coupling strengths
\begin{equation}
\begin{split}
J_1^{(eval)}=1.39998,&\hskip0.25cmJ_2^{(eval)}=1.48005,\\
J_3^{(eval)}=1.06003,&\hskip0.25cmJ_4^{(eval)}=0.800058,\\
J_5^{(eval)}=1.36050,&\hskip0.25cmJ_6^{(eval)}=0.970524,\\
J_7^{(eval)}=&\,0.660894,
\end{split}
\end{equation}
which are a very accurate estimation of the original ones.

\section{Remarks}
\label{remarks}
We have presented the working features of a scheme for Hamiltonian tomography that allows the identification of coupling parameters in various classes of spin-chain models through the study of the time dynamics of a single spin. The method is designed to work in a scenario of restricted access to  the components of the chain. As no initial state preparation is necessary, one can perform measurements by interspersing them with the natural evolution of the system. Besides data acquisition, only a simple post-processing step is necessary: no conservation law associated with the interaction or {\it a priori} knowledge on the state of the system is required. Even when a spin-chain is affected by environmental influences, the Hamiltonian tomography remains possible and reliable. Given the crucial role that proper coupling patterns have in the interference effects behind quantum many-body phenomena, non-demanding diagnostic methods, like the one proposed here, are important tools which need to be developed.

\section*{Acknowledgments}
We thank N. Lo Gullo for discussions. We acknowledge support from the UK EPSRC. C.D.F. is supported by the Irish Research Council for Science, Engineering and Technology. M.P. thanks the UK EPSRC (EP/G004579/1) for financial support.

\end{document}